# A fast method to simulate travelling waves in nonhomogeneous chemical or biological media

Kristóf Kály-Kullai

Institute of Physics, Department of Chemical Physics, Budapest University of Technology and Economics, Budapest H-1521, Hungary, E-mail: kk222@ural2.hszk.bme.hu

**Abstract**

Waves in excitable media can be treated by a simple geometric theory. The propagation velocity is assumed known and evolution of wave fronts is determined by elementary physical principles (Fermat's principle, Huygens' principle). Based on this geometric theory a fast computational method is developed. By this method the distorting effect of the spatial grid is avoided. The method is applied to the cases when a circular obstacle is surrounded by a homogeneous and heterogeneous medium, respectively. The numerical simulations show that the method is convenient, fast and reliable.

## 1. Introduction

Wave phenomena are widely studied and applied in different fields of natural sciences. The definition of wave is not uniformly accepted: there are various approaches (propagation of disturbances, propagation of energy, periodic spatio-temporal processes, solutions of the wave equation, etc.). Here we will use the term "wave" for a spatio-temporal process described by a function of the form

$$u(t, \mathbf{r}) = A(\mathbf{r})f(t - S(\mathbf{r})). \qquad (1)$$

Here t is the time, **r** is the position vector, A is the amplitude, f is the phase, and S is the eikonal. This terminology is in accordance with that one commonly used in the special case of harmonic waves (when f is sinusoidal function of its argument).

The form (1) describes an undistorted propagation of a signal. If A is not constant, then the signal is simply attenuated. The eikonal S measures the time needed for the propagation. Thus the time dependence will be essentially the same at every point **r**, except of the attenuation (characterised by A) and the time delay (measured by S).

Wave fronts are the equieikonal surfaces:

S(**r**) = constant.

Travelling waves are represented by propagating fronts. That means a continuous transformation of the wave fronts into themselves. Evolving fronts or in other words "moving interfaces" occurs in very wide variety of applications. Fast Marching Method, introduced by Sethian (1996) and Level Set Method, introduced by Osher and Sethian (1988) are numerical techniques designed to track the evolution of interfaces. For details of these methods, and the possible fields of applications (geometry, fluid dynamics, shape recovery in image processing, semiconductor manufacturing, and so on) see the book [12].

Wave functions (1) are often presented as solutions of partial differential equations. In our case chemical waves are solutions of certain reaction-diffusion equations in an "excitable" medium [4, 11, 5].

There are two traditional approaches for computer simulations of chemical waves:

i) application of cellular-automaton models [8],

ii) numerical solution of the reaction-diffusion equations [2].

A common problem of the above approaches is the spatial grid. The actual front is the set of points in a special state, and it is difficult to eliminate the distorting effect of the grid. A further problem is that numerical solving of the reaction-diffusion equations is time-consuming. Our new method is based on the geometric theory of waves. It does not require the numerical solution of any differential equation and hence it enables us to construct a fast method.

## 2. The geometric theory

In many simple cases of wave propagation the amplitude has no significant role, it can be considered constant. The process of propagation can be treated without involving the amplitude. In the geometric theory of waves we define rays and fronts, we study the evolution of fronts provided that the wave speed v is given as a function of position vector **r** [1, 14]. The geometric theory is based on Fermat's principle of the least propagation time.

Let us take two points in the excitable medium, and let g be a curve with the endpoints A and

B. The propagation time belonging to g will be denoted by t(g(A, B)), that is

$$t(g(A,B)) = \int_g \frac{1}{v(\mathbf{r})} ds.$$

A curve is called *Fermat ray* or extremal if there is no other curve with same endpoints having less propagation time. This Fermat ray belongs to a pair of points, namely the endpoints. Now we generalise the concept of Fermat ray for the case when the initial point is not fixed, but it is an arbitrary point of a given initial set H. The Fermat ray between the initial set H and an endpoint B is defined as follows. Consider all the propagation times of the Fermat rays belonging to varying A and fixed B, where A ∈ H, and select the one to which the propagation time is minimal [14]. We refer to it as the Fermat ray between H and B. In generic case, there is a unique Fermat ray between a point pair or between a set and a point. However, there may be exceptional cases, when several different Fermat rays belong to a point pair or a given set and a point (singularity theory: [1]).

Now let H be the initial front $F_0$, and let the endpoint B be arbitrary. This way a family of Fermat rays is generated. The subsequent wave fronts are the orthogonal trajectories of that family of Fermat rays starting from the initial front $F_0$.

Sieniutycz elaborated the exact formulation of the geometric wave theory in frame of variational calculus for inhomogeneous and anisotropic medium when the velocity depends on space and direction [13].

Sainhas and Dilao numerically studied the solutions of reaction-diffusion equation for the Brusselator model and they deduced the validity of the elementary laws of geometrical optics [10]. His results showed that the reaction-diffusion theory was in agreement with the geometrical wave theory in that special case.

The first application of the geometrical approach to biology goes back to Wiener and Rosenblueth [19]. They constructed a model to the propagation of excitations in nerve system and cardiac muscle.

These waves of excitations in biology are very similar to the chemical waves. A simple analogue, which illustrates the most striking character of chemical waves, is the process of prairie fire. Starting from an initial burning front, the process is easily predicted if the velocity of propagation is known.

Our aim was to develop a fast computer program to simulate travelling waves in excitable media based on the geometric theory. The main goal was to eliminate the distorting effect of the spatial grid.

## 3. Description and application of the method

### 3. 1. Modelling of the active medium, front and obstacle

In this model the unit of time is step number, and the unit of length is pixel. The excitable media is modelled with a 2 dimensional (x-y) lattice (matrix) called *prairie*, where these bytes represent the state of the excitable media in that point, and when drawing on the screen one byte determines the colour of one pixel. A point of the media can have three possible states:

- *Obstacle state*: the wave can not pass through these points, and they keep their state during the whole simulation.

- *Resting state*: the wave can pass through these points and after that the state of point changes to refractory state. (Evoking the prairie fire analogue, this operation is called burning. However, the duration of the burning is regarded to be zero here, there is no "burning" state in our model.)

- *Refractory state*: the wave cannot pass through these points. While burning, a point in resting state changes to refractory state, and after time T (recovery time) the point will change back to resting state again. The latter stage of the process is called recovering according to the prairie fire terminology.

The recovery algorithm needs the last time (number of steps) when a point turned to the refractory state. These times are stored in an other matrix, called *last time matrix*. This matrix is indexed in the same way as the prairie matrix.

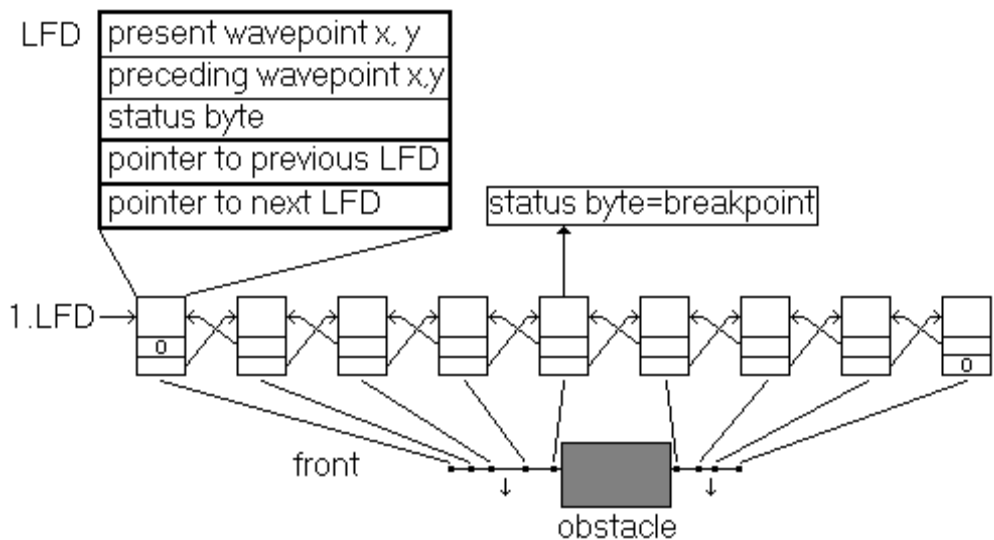

Fig. 1. The LFD structure and the front modelled as a linked list of LFDs. This figure shows a front around an obstacle.

In our model the wave front is not the set of points in a special state, it is stored separately from the prairie, and it is a linked set of local front determinators. A *local front determinator* (LFD) is a structure, which contains the coordinates of the wavepoint in the present and in the preceding step (here we use the terminology: *point* means a point of excitable media, and *wavepoint* means a point of front), pointers to the previous and next LFD, and a status byte indicates whether the LFD is on the end of one continuous part of the front (it is *breakpoint*) or not. The front is an ordered, linked list of LFDs (see Fig. 1.), so they determinate the actual shape of the front. The distance of their present coordinates is kept between two given values. It is important that the coordinates of the wavepoints are floating point numbers. This way we can eliminate the spatial grid together with its distorting effect from the front calculations. Rounding is necessary only when the front is drawn on the screen (see Fig. 2.).

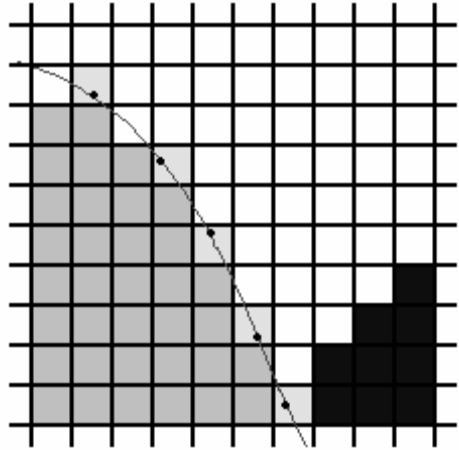

Fig. 2. The active medium and the wave front. One square represents one point of the prairie. The white squares are in resting state, the dark grey squares are in refractory state. The wavepoints (black points) are drawn to the light grey squares. The black squares are in obstacle state.

Applying the Fermat's principle one wavepoint moves perpendicularly to the front with a speed determined by the velocity field (a given v(x,y) function). The orientation of the front is defined by the list. This orientation selects between the two possible directions of wave propagation in such a way that the direction of the wave propagation and the direction of the orientation form a right hand system. Inhomogeneous media can be described with a non-constant velocity field. (The model can be applied for anisotropic media as well but here we regard the isotropic case only.)

### 3. 2. Start of the simulation

### 3. 2. 1. Back to the default state

Before a new simulation is started the previous wave fronts (if there was any) is deleted from the memory. Then the prairie is filled with resting state points, except for the edges of the prairie, where one pixel wide border is created from obstacle points. This way any previous obstacle is also deleted. The last time matrix is filled with zeros, and the step number is also zero.

### 3. 2. 2. Preparing the new obstacles and an initial front

The program allows us to change the velocity field (v(x,y) function).

The border curve of an obstacle can be given by two functions $y_i(x)$ (i=1,2) in the region [$x_1$,

$x_2$]. The points of the prairie between these two boundary curves will be in obstacle state.

For convenient the actual initial front is given here by a function x(y) given on a certain interval [$y_1$, $y_2$]. In this interval new LFDs are created. The present y coordinates of these wavepoints are stored with y started from the low limit and stepping by one pixel, and the x coordinates are stored with the x(y) values. The preceding coordinates also assume these values. After that the front is ordered (see 3.3.4.).

### 3. 3. One step of the simulation

#### 3. 3. 1. Overview

In one step the following operations should be performed:

1) Recovery: any point, which has spent T (recovery time) in the refractory state, changes its state to resting.

2) Moving of the front: going along the linked list every LFD is moved perpendicularly to the front.

3) Ordering the new front.

4) Quadrangle burning: the points are changed from resting state to refractory one in a quadrangle determined by two neighbouring LFDs present and preceding coordinates.

5) Circle burning: the points change their state to refractory one in a circular range around a wavepoint at the breakpoints of the front.

6) Drawing.

7) Increasing the step number.

#### 3. 3. 2. Recovery

The state of a point is characterised by a number: 1 represents the resting state and 0 the refractory one. The program checks all of the prairie's points, and if one point is found in refractory state, then the new state of the point is $\Theta(t-t_0-T)$, where T is the *recovery time* (it can be given by the user), t is the actual step time, and $t_0$ is picked up from the last time matrix, which stores the step number when the point changed its state to refractory, and $\Theta$ is the unit step function:

$$\Theta(x) = \begin{cases} 1, & \text{if } x \geq 0, \\ 0, & \text{if } x < 0. \end{cases}$$

This method can be generalised allowing additional possible states between the refractory and resting state. Then instead of Θ another recovery function should be used. Wiener and Rosenblueth [19] used the concept of *epoch number* to characterise the stages of the refractory state.

### 3. 3. 3. Moving of a LFD

The LFD is moved perpendicular to the front with an extent determined by the velocity field (a given v(x,y) function). Calculating of the perpendicular direction needs the derivative of the front in the actual point. To calculate it, quadratic Lagrange interpolation is done to the actual, previous and next LFDs. If the actual LFD is at one end of a continuous part of the front, then interpolation is done to the actual, and the two previous (or next) LFDs. If a continuous part contains only two LFDs, then linear interpolation is applied, and if it contains only one LFD, then it moves horizontally. To increase the accuracy, if the x-projection of the interval containing these points is bigger than the y-projection, then the interpolation is done as an y(x) function, otherwise as an x(y) function.

From the coefficients of the interpolation polynomial the approximating value of the derivative in the actual wavepoint is calculated. From that, the perpendicular unit vector (direction vector, called **i**) is calculated by using the right hand convention (in 3.1.).

Then the coordinates of the present wavepoint are copied to the coordinates of the preceding wavepoint. After that, the direction vector is multiplied by the extent of the moving (see later) and then it is added to the coordinates of the present wavepoint to produce its new values.

Because of this wavepoint changing method, when interpolation is calculated, the coordinates of previous wavepoints are picked up from preceding wavepoint of the previous LFD, and the coordinates of the actual and next wavepoints are picked up from the present wavepoints of their LFDs (see Fig. 3.).

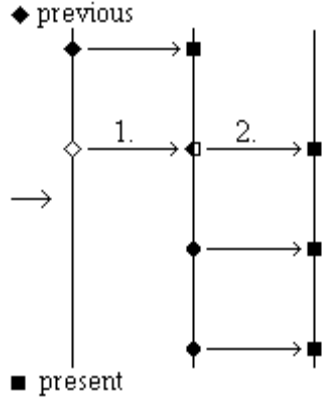

Fig. 3. Moving of LFDs. 1.: first step, 2.: second step.

To calculate the progress of one wavepoint, first the value of the velocity function is evaluated in the present wavepoint of the actual LFD, and is compared with the velocity in the point after the calculated progress. If these values are not equal, or the option "always fine moving" is active then progress is calculated as an integral. (The "always fine moving" option is useful for example in the case when the velocity periodically changes.) Then the new coordinates are

$$x_k(t) = x_k(t_0) + \mathbf{i}_k \int_{t_0}^{t} v(x_k(t')) dt',$$

where k is the index of the vector components (k=1,2 x and y components), **i** is the direction vector, t0 is the actual time and t is the time of the new wavepoint, in this model t=t0+1. The above expression should be approximated with the sum:

$$x_k(t) = x_k(t_0) + \mathbf{i}_k \sum_{j=1}^{n} v(x_k(t_0 + (j-1)\Delta t)) \Delta t,$$

where $\Delta t = 1/n$, and n is a variable parameter called *fine step number*.

Having calculated the new wavepoint, the pathway of propagation is checked; whether there is a point on it in refractory or in obstacle state. If the answer is yes, then the LFD stay on the last point in resting state and its status byte changes to breakpoint status.

### 3. 3. 4. Ordering

The aim of ordering is to keep the density of LFDs in the neighbourhood of a given value in the case of expanding or contracting fronts. An additional aim is to delete LFDs if neighbouring ones are in the breakpoint status. This occurs when the front collides to an obstacle. In this case the colliding points are marked as breakpoint, and they are deleted from the list except the first one. The remaining LFD in breakpoint status marks that the front is broken between the breakpoint and the next LFDs.

To execute the ordering, the program goes through the LFDs, and with each one the following checks and modifications are done:

1) If both actual and next LFDs are in breakpoint status, then the actual LFD is deleted, and the check continues with the next LFD.

2) While the distance of actual and next LFD is bigger than a prescribed value (called *maxdist*), new LFD is created between them. To create it quadratic Lagrange interpolation is used in the same manner as described in 3.3.3. The coordinates of the new wavepoint are created so, that the independent variable of the interpolation is the mean of the old wavepoints same coordinates, and on other coordinate is the value of the interpolation polynomial on the previous place.

3) While the distance between the actual and next LFD is smaller than a prescribed value (called *mindist*), and the distance between the actual LFD and the one after the next is smaller than the value of maxdist, the next LFD is deleted.

4) If the actual LFD is in breakpoint status, and it is nearer to the next LFD as a prescribed minimum distance, then the actual point will be no more in breakpoint status. (The two front parts grow together.)

### 3. 3. 5. Quadrangle burning

The quadrangle burning changes the points on the area where the front passed through, from resting state to refractory. To realise this, the program goes through the LFDs, and the points in resting state change their state to refractory one in a quadrangle determined by the present and preceding coordinates of the actual and the next LFDs.

### 3. 3. 6. Circle burning

Circle burning yields circular front parts at the breakpoints. In homogeneous media these front parts are exact circle arcs, but in nonhomogeneous media they are not necessarily. This procedure makes it possible that two front parts, which are separated by the obstacle, reunite again after passing by the obstacle.

The coordinates of new LFDs are calculated by rotating the present coordinates of the breakpoint around its preceding coordinates. The rotation angle φ is calculated from the formula:

$$\varphi = \begin{cases} \dfrac{\min\text{dist} + \max\text{dist}}{2r} & \text{, if it is smaller than } \pi/2 \\ \pi/2 & \text{, otherwise} \end{cases}$$

where r is the distance between the present and preceding wavepoints of the breakpoint, mindist and maxdist are defined in 3.3.4. In homogeneous medium r is the radius of the circle arc on which the present wavepoints of the new LFDs are.

We repeat the rotation until n times, where n=int(2π/φ). After a rotation the preceding wavepoint of the new LFD will be the same as the breakpoints one. To calculate the present wavepoint, a modified version of LFD moving (3.3.3.) is used. Here the direction (**i** vector) is calculated by rotating the present wavepoint of the breakpoint around the preceding wavepoint of it with an angle extent jφ, where j is the number of rotations (j=1..n). The sign of the angle depends on the location of the breakpoint. If it is in a beginning of a continuous front part, the angle is negative (clockwise), and if it is in the end of it, the angle is positive (anti-clockwise). Having the **i** vector, the moving continues as described in 3.3.3., so this method works well also in nonhomogeneous media. (See Fig. 4.)

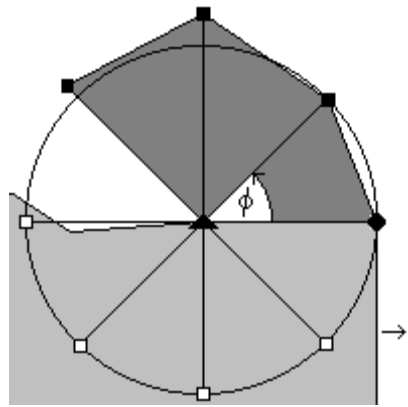

Fig. 4. Circle burning. In this example the angle φ is positive. The small triangle at the centre is the preceding wavepoint of the breakpoint, the small circle is its present wavepoint, the squares are the new LFDs, but only the filled ones stay in the list, the empty ones are deleted in the procedure of moving.

After the new LFDs were created, the program goes through the new LFDs, and the points in resting state transform into refractory state in a triangle determined by the present and the common preceding coordinates of the actual and the previous LFDs respectively.

Circle burning is done only after the second step of simulation in order to prevent propagation backward. After the second step the set of the points in refractory state behind the front is thick enough to prevent backward propagation.

### 3. 3. 7. Drawing

The program goes through the points of the prairie to check their states. The colour of the corresponding pixel is determined by the state of the point.

Next, the program goes through the LFDs, round their present coordinates, and changes the colour of the corresponding pixel red. The (i,j) pixel is the corresponding pixel of any wavepoint in the interval [i-0.5,i+0.5),[j-0.5,j+0.5).

## 4. Results and discussion

### 4. 1. Computations

We have simulated chemical waves in uniform and nonuniform membranes rotating around an obstacle (a hole in the membrane). Theoretical and experimental studies on such waves have already been published [7, 14, 6, 16, 17]. The experimental wave fronts show a good agreement with the theoretical predictions based on the geometric theory of waves: the wave fronts were involutes of the obstacle [7, 6] (Fig. 5. shows involutes of a circle). Remark that the curvature effect [3, 15] was negligible in those experiments.

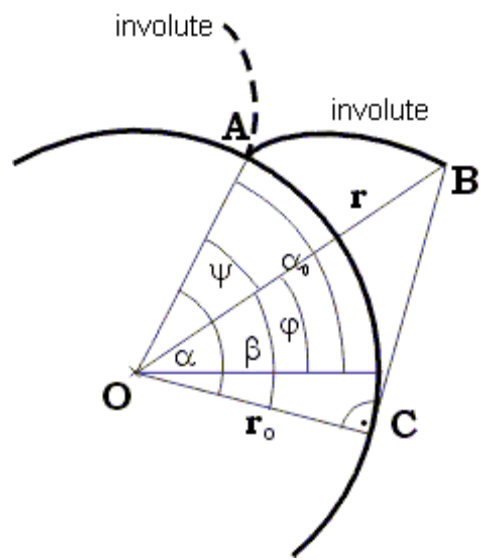

Fig. 5. Involutes of a circle.

Fig. 6 compares the experimental results [18] with the present simulations for waves rotating around a circular obstacle. The numerical parameter values are: mindist=$\sqrt{0.9}$, maxdist=$\sqrt{2.5}$ (see in 3.3.4.) and the fine step number (see in 3.3.3.) n=25 in all simulations. Some physical parameters, and their values were varied, such as the T recovery time, $R_0$ obstacle radius, $R_l$ limit radius (it is the radius of the membrane), v velocity of propagation (in homogeneous media). In heterogeneous media there are some additional physical parameters: $R_i$ interface radius (this interface separates the media an inner slower annular region and an outer faster one), $v_s$ velocity in the inner slower region, $v_f$ velocity in the outer faster region and d distance between the centres of the obstacle and the interface, respectively.

For Fig. 6, the parameter values are:

a) Homogeneous medium: T=25, $R_0$=20, $R_l$=100, v=0.5

b) Heterogeneous medium, symmetric arrangement, minimal loop (the closed path requiering shortest circumference time around the obstacle, see [16, 17]) is the border of obstacle: T=20, $R_0$=25, $R_i$=75, $R_l$=125, $v_s$=0.5, $v_f$= 1.0, d=0

c) Heterogeneous medium, symmetric arrangement, minimal loop is the interface: T=20, $R_0$=25, $R_i$=75, $R_l$=125, $v_s$=0.5, $v_f$= 2.0, d=0

d) Heterogeneous medium, asymmetric arrangement, minimal loop is the obstacle: T=20, $R_0$=25, Ri=75, Rl=125, vs=0.5, $v_f$= 1.25, d=35

e) Heterogeneous medium, asymmetric arrangement, minimal loop is the interface: T=20, R0=25, $R_i$=75, $R_l$=125, $v_s$=0.5, $v_f$= 1.7, d=35.

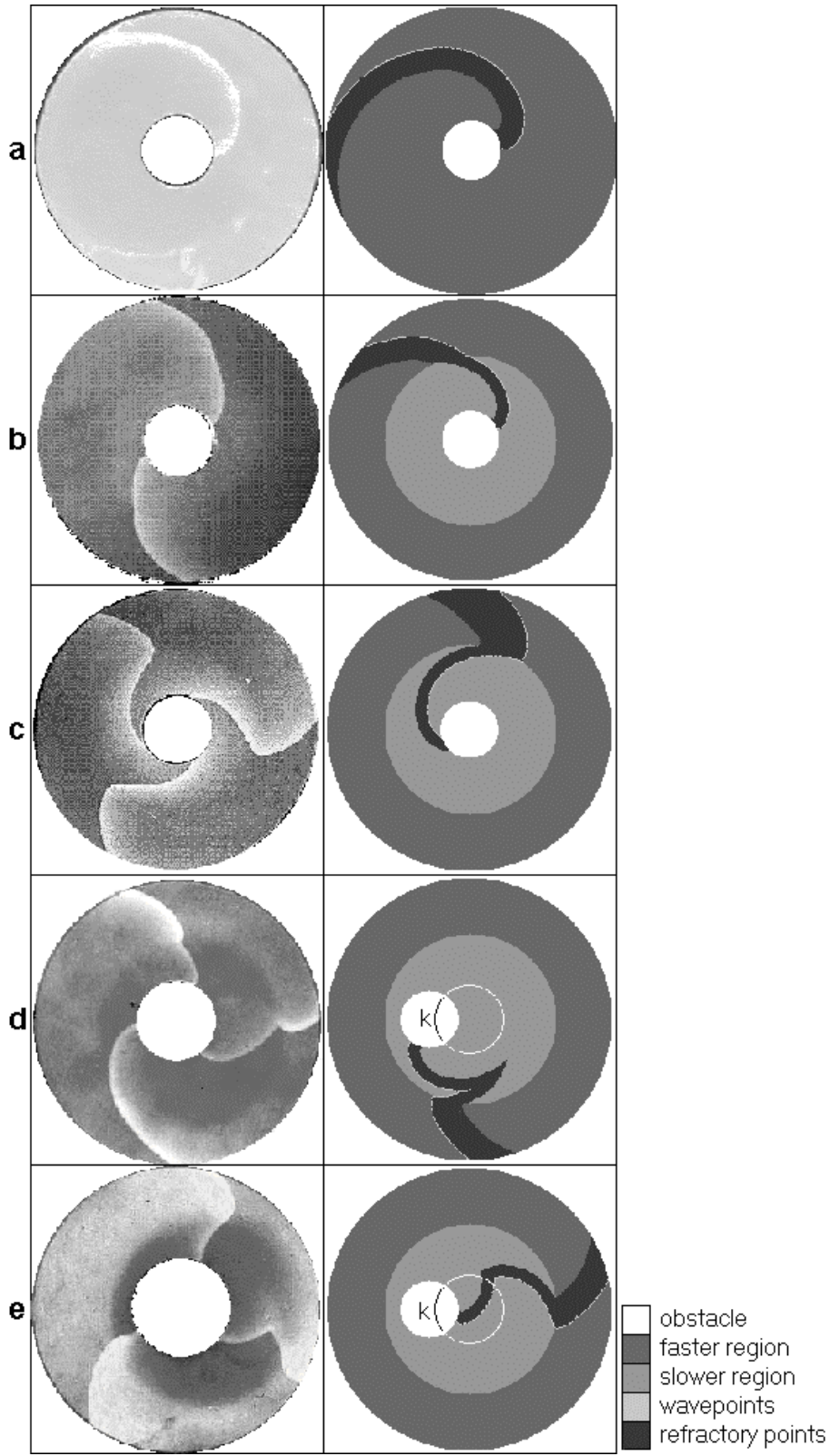

Fig. 6. The experimental (on the left column) and simulated (right column) results:

a) Homogeneous medium, b) Heterogeneous medium, symmetric arrangement, minimal loop is the obstacle, c) Heterogeneous medium, symmetric arrangement, minimal loop is the interface, d) Heterogeneous medium, asymmetric arrangement, minimal loop is the obstacle, e) Heterogeneous medium, asymmetric arrangement, minimal loop is the interface. The breakpoint in the front moves on the interface. However, in case of asymmetric arrangement (d, e) there is a second breakpoint, which moves partly on the circle k inside of the inner region. This circle is called caustic, and in our case it is the envelope of the rays departing inward from the interface with the critical angle of the total reflection. For details, see[9].

## 4. 2. Comparison with analytical predictions

In the case of homogeneous annular media, we made some quantitative study. In this case the wave front is an involute of a circle, and its equation is known in polar coordinates:

$$\varphi(r) = \alpha_0 \pm \left[ \sqrt{\left(\frac{r}{R_0}\right)^2 - 1} - \arccos\left(\frac{R_0}{r}\right) \right], \quad (2)$$

where $R_0$ is the radius of the circle and $\alpha_0$ is the angle where the involute starts (see Fig. 5).

We can fit an involute to the present coordinates of the LFDs. To fit it, we minimise the khi square function:

$$\chi^2(r_0, \alpha_0) = \sum_{i=1}^{n} \frac{(\varphi_i - \varphi(r_i))^2}{N},$$

where $r_i$ and $\varphi_i$ are the present polar coordinates of the LFDs, N is the number of LFDs and $\varphi(r_i)$ is calculated from the equation (2). To minimise it, we use the Nelder-Mead simplex method [9].

First, the fixed parameters were: $R_l$=125 pixel, v=1 pixel/step number. The varying parameter was $R_{0,geo}$ (it is the given $R_0$ value in the simulation). The differences between the $R_{0,fit}$ fitted radius and the given $R_{0,geo}$ radius ($\Delta R_0$) are shown in Fig. 7. It is seen that $\Delta R_0$ is nearly constant.

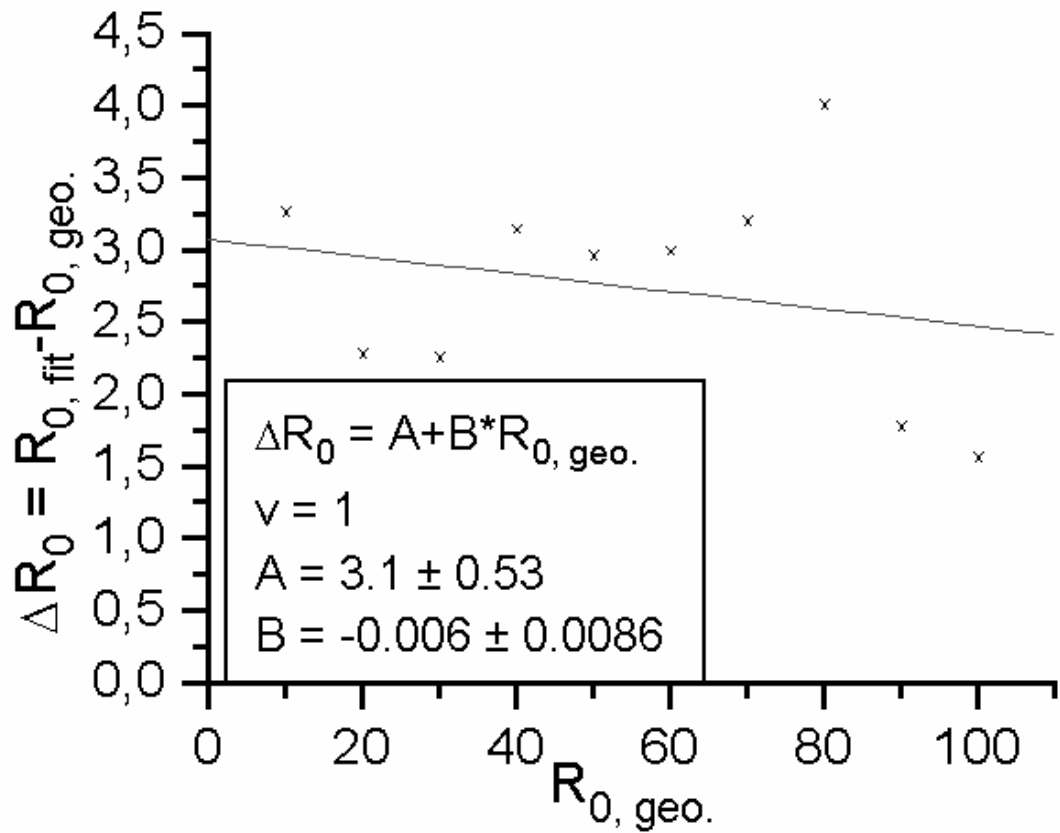

Fig. 7. The difference between the given radius of the obstacle and the radius fitted to the calculated front ($\Delta R_0$) vs. the radius of the obstacle ($R_{0,geo}$).

Fig. 8 shows $R_0$ vs. v, when $R_0$=20 pixel. $\Delta R_0$ is turned to be nearly a linear function of the velocity v.

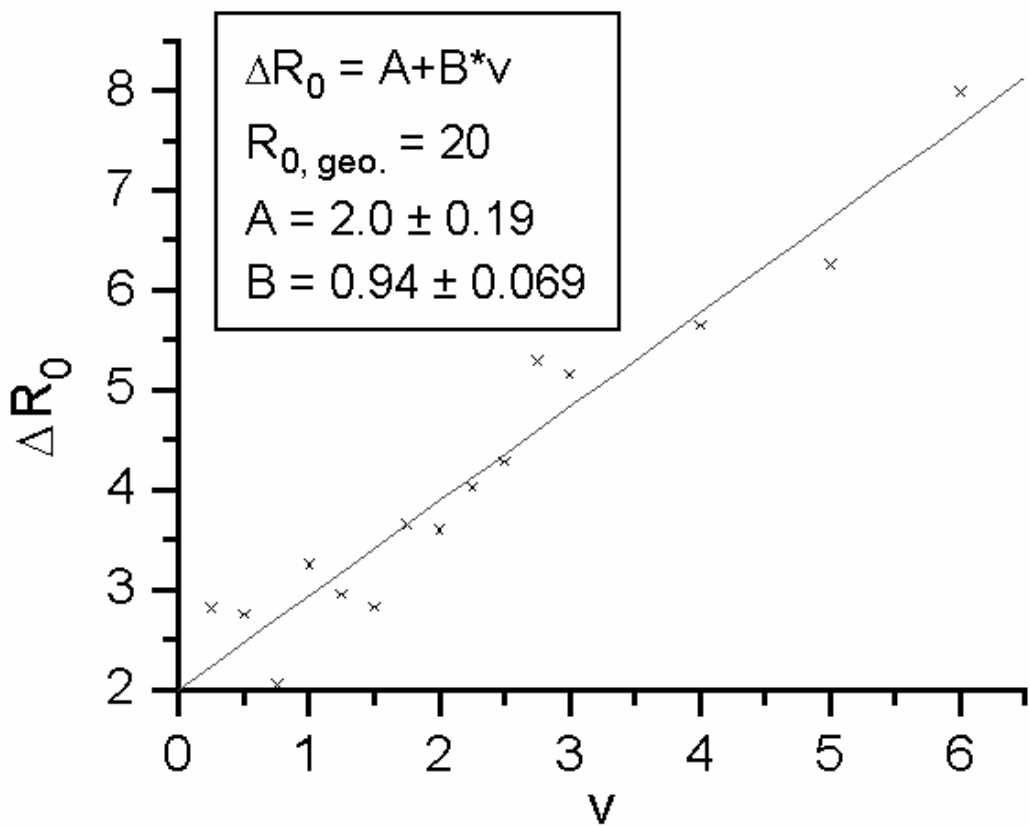

Fig. 8. The difference between the given radius of the obstacle and the radius fitted to the calculated front ($\Delta R_0$) vs. the velocity (v).

These studies prove that if the numerical velocity is smaller (the time discretisation is finer) or the obstacle radius is bigger (the spatial discretisation is finer) then the relative error of the simulation decreases.

All these results show that the present fast computer simulation method based on the geometric theory of waves is a suitable tool for numerical study of chemical waves.

## Acknowledgements

I wish to thank Henrik Farkas, Zoltán Noszticzius and András Volford for their helpful discussion. This work was partially supported by OTKA (T-30110) grant.